\documentclass{JHEP3} 
\JHEPspecialurl{http://jhep.sissa.it/JOURNAL/JHEP3.tar.gz}
\pdfoutput=1

\usepackage{subfigure}
\usepackage{rotating}
\usepackage{amssymb}
\usepackage{amsmath}
\usepackage{multirow}
\usepackage{appendix}

\newcommand{\lsim}{\mathrel{\mathop{\kern 0pt \rlap
  {\raise.2ex\hbox{$<$}}}
  \lower.9ex\hbox{\kern-.190em $\sim$}}}
\newcommand{\gsim}{\mathrel{\mathop{\kern 0pt \rlap
  {\raise.2ex\hbox{$>$}}}
  \lower.9ex\hbox{\kern-.190em $\sim$}}}

\newcommand{\be}{\begin{equation}}
\newcommand{\ee}{\end{equation}}
\newcommand{\bea}{\begin{eqnarray}}
\newcommand{\eea}{\end{eqnarray}}

\def\ptmiss{\not\!\!{p_T}}

\newcommand{\ntrl}[1]{\chi^0_#1}

\title{Lepton flavour violating signature in supersymmetric $U(1)^\prime$ seesaw models at the LHC }

\author{Priyotosh Bandyopadhyay$^{a,b,1}$, Eung Jin Chun$^{c,1}$\\

 $^{a}$Department of Physics, University of Helsinki and
    Helsinki Institute of Physics, 
  FIN-00014, Helsinki, Finland\\
$^b$Dipartimento di Matematica e Fisica "Ennio De Giorgi", Universit`a del Salento and INFN, Via Arnesano, 73100, Lecce, Italy\\
$^c$Korea Institute for Advanced Study,
Seoul 130-722, Korea\\
Email:  \email{$^1$priyotosh.bandyopadhyay@helsinki.fi,
  priyotosh.bandyopadhyay@le.infn.it, $^2$ ejchun@kias.re.kr} }

\preprint{KIAS-P14071/HIP-2014-23/TH}

\abstract{We consider a $U(1)'$ supersymmetric seesaw model in which  a right-handed sneutrino is a thermal dark matter candidate 
whose relic density can be in the right range due to its coupling to relatively light $\tilde{Z'}$, the superpartner of the extra gauge boson $Z'$. Such light $\tilde Z'$ can be produced at the LHC through cascade decays of colored superparticles, in particular, stops and sbottoms, and then decay to a right-handed neutrino and a sneutrino dark matter, which leads to lepton flavor violating signals of same/opposite-sign dileptons (or multileptons) accompanied by large missing energy.
Taking some benchmark points, we  analyze the opposite-  and same-sign dilepton signatures and the corresponding flavour difference i.e., ($2e-2\mu$).  It is shown that $5\sigma$ signal significance can be reached for some benchmark points with very early data of $\sim 2$ fb$^{-1}$ integrated luminosity. In addition, $3\ell$ and $4\ell$ signatures also look promising to check the consistency in the 
model prediction, and it is possible to reconstruct the $\tilde{Z'}$ mass 
from $jj\ell$ invariant mass distribution.}

\begin{document}


\section{Introduction}

Among various reasons for requiring theories beyond the Standard Model (SM), experimental evidences for tiny neutrino masses and dark matter, and a theoretical requirement for naturalness of the electroweak scale would be key elements and related to each other. 
 The first candidate model addressing these features would be Minimal Supersymmetric Standard Model (MSSM) where the neutrino masses and mixing are explained by R-parity and lepton-number violation  \cite{rpv} and the dark matter consists of a slowly decaying gravitino as the lightest supersymmetric particle (LSP). In this paper, we wish to explore another possibility which can provide richer collider signatures.
A sumpersymmetric seesaw model \cite{Mohapatra05} associated with an additional gauge symmetry $U(1)'$ \cite{Langacker08} would be one of such examples.

In supersymmetric theories with R-parity, the LSP is stable and thus a neutral LSP, typically a linear combination of neutral gauginos and Higgsinos,  becomes a good thermal dark matter (DM) candidate if supersymmetry (SUSY) is broken around the TeV scale \cite{Jungman95}.
The SUSY breaking can radiatively induce the $U(1)'$ breaking in addition to the usual electroweak symmetry breaking, which determines the seesaw scale also at ${\cal O}$(TeV) \cite{Khalil07}.
 A typical example of the additional abelian gauge symmetry $U(1)'$ is $U(1)_{B-L}$  which requires the presence of right-handed neutrinos (RHNs) for the anomaly-free condition, and thus realizes the TeV-scale seesaw mechanism. Let us note that we do not assume any grand unification theory as the origin of our 
model and the grand unification structure will be used for a convenient guide to a theoretically consistent 
model gauranteeing the anomaly free condition, and so on.
In our framework, a right-handed sneutrino (RHsN) can be  the LSP and thus another good  dark matter candidate whose thermal relic density is in the right range if the $U(1)'$ gaugino $\tilde Z'$, the superpartner of 
the $U(1)'$ gauge boson $Z'$, is relatively light \cite{Bandyo11}.

In this paper, we analyze LHC signatures of the extra gaugino $\tilde Z'$ which can be pairly produced mainly through third generation
squark cascades and then decay to a RHN and a RHsN DM. Due to the Majorana nature of a RHN, $N$,  it can decay to both-sign leptons, 
$N \to l^\pm W^\mp$, leading to the lepton number violating signature of same-sign dilepton (SSD) events 
in addition to the usual opposite-sign dilepton (OSD) events \cite{Keung83}. 
Furthermore, Yukawa couplings of a RHN are generically flavour-dependent and thus lead to lepton flavour violating decays, e.g., 
$\mbox{Br}(N \to e^\pm W^\mp) \gg \mbox{Br}(N \to \mu^\pm W^\mp)$.   
Taking some benchmark points, we carry out a Pythia-FastJet level collider simulation at the 14 TeV LHC for multilepton final states to  study the prospect for detecting such lepton flavour violating signatures manifested by a flavour difference, e.g., 
$2e - 2 \mu$, in both same-sign and opposite-sign dilepton  final states.  
Similar phenomenon should appear also in multi-lepton channels such as a tri-lepton difference $3e-3\mu$.
As we will see,  $5\sigma$ discovery is expected in the $2e-2\mu$ SSD final states for an optimistic benchmark point 
even with $\sim 2$ fb$^{-1}$ integrated luminosity at the very early stage of LHC14. It is straightforward to follow the same procedure for the opposite case of $\mbox{Br}(N \to \mu^\pm W^\mp) \gg \mbox{Br}(N \to e^\pm W^\mp)$ resulting in a similar conclusion.

Let us here recall that the neutrino mass models with R-parity and lepton number violating couplings lead to similar signatures \cite{rpv2}. In particular, there could appear an interesting connection of the electron excess assumed in this work and  the neutrinoless double-beta decay caused by R-parity violation as pointed out by Allanach et.al.~\cite{allanach}.  
Furthermore, the supersymmetric left-right symmetric model \cite{nemevsek} can also lead to the similar signatures except a possible $W'$  appearance in the next LHC run.

This paper is organized as follows. In Section 2, we describe a general $\tilde Z'$ phenomenology 
by taking a specific $U(1)'$ model. Considering all the relevant experimental constraints 
on SUSY parameter space, three benchmark points are set up in Section 3, and corresponding production 
rates and decay branching fractions of squarks are calculated in Section 4. A detailed LHC phenomenology 
of lepton number and flavour violating signatures are analyzed in Section 5, and we conclude in Section 6.

\section{$\tilde{Z}^{\prime}$ Phenomenology}

Among various possibilities of an extra gauge symmetry $U(1)'$ and
the presence of the associated right-handed neutrinos
\cite{Langacker08}, we will take the $U(1)_\chi$ model for our
explicit analysis as in \cite{Bandyo11}. The particle content of our $U(1)_\chi$ model
is as follows:
\begin{equation}\label{chargeqn}
\begin{array}{c|ccc|cc|c|cc}
SU(5) & 10_F & \bar{5}_F & 1 (N) & 5_H & \bar{5}_H & 1 (X) & 1
(S_1) & 1 (S_2) \cr \hline 2 \sqrt{10} Q' & -1 & 3 & -5 & 2 & -2 &
0 & 10 & -10 \cr
\end{array}
\end{equation}
where $SU(5)$ representations and $U(1)'$ charges of the SM
fermions ($10_F, \bar{5}_F$),  Higgs bosons ($5_H,
\bar{5}_H$), and additional singlet fields ($N, X, S_{1,2}$) are
shown. Here $N$ denotes the right-handed neutrino, $X$ is an
additional singlet field fit into the 27 representation of $E_6$,
and we introduced more singlets $S_{1,2}$, vector-like under
$U(1)'$, to break $U(1)'$ and generate the Majorana mass term of
$N$ \cite{Khalil07}. Note that the right-handed neutrinos carry the largest charge under
$U(1)'$ and thus the corresponding $Z'$ decays dominantly to
right-handed neutrinos. Furthermore, the additional singlet field
$X$ is neutral under $U(1)'$ so that it can be used to
generate a mass to the $U(1)'$ Higgsinos as will be discussed
below.

The gauge invariant superpotential in the seesaw sector is given by
\begin{equation}\label{wpot}
W_{seesaw} = y_{ij} L_i H_u N_j + {\lambda_{N_i}\over2} S_1 N_i
N_i\;,
\end{equation}
where $L_i$ and $H_u$ denote the lepton and Higgs doublet
superfields, respectively. After the $U(1)'$ breaking by the
vacuum expectation value $\langle S_1 \rangle$, the right-handed
neutrinos obtain the mass $m_{N_i}=\lambda_{N_i} \langle S_1
\rangle$ and induce the seesaw mass for the light neutrinos:
\begin{equation}
 \widetilde{m}^\nu_{ij} = - y_{ik} y_{jk}{ \langle H_u^0\rangle^2 \over
 m_{N_k} }\,.
\end{equation}

In this type of models the RHN can be produced from the $Z'$ decay, $Z' \to NN$, and 
then can go through a flavour violating decay, $N\to e W$.
The recent mass bound of $Z'$ from LHC, pushes $m_{Z'}\gsim 2\sim3$ TeV depending on $U(1)'$ types \cite{z'bound}.
This reduces drastically the production cross-section $pp\to Z' \to NN$ as shown in \cite{mdl}. 
The other source of RHNs could be via the production  of  $\tilde{Z'}$. 
Once produced, it can have a decay to a RHN ($N$) and a RHsN LSP ($\tilde N_1$), 
i.e., $\tilde Z' \to N\tilde{N_1}$. 

As shown in  \cite{mdl}, the direct production cross-section of  $pp \to \tilde{Z'}\tilde{Z'} $ is not encouraging 
 even for 14 TeV at the LHC.  Another supersymmetric source could be the cascade decays of the strongly interacting superpartners, i.e., squarks and gluinos.  Similarly to the cascade decays of electroweak gauginos and higgsinos that can produce gauge bosons and Higgs 
in the minimal supersymmetric standard model \cite{susycd1,susycd2}, one can  have extra gauginos, $\tilde{Z'}$, 
 produced via cascade production of squarks and gluinos in a supersymmetric $U(1)'$ model. 
From \cite{mdl} (Figure 13), we can see that the right-handed down type squark mostly decays to $\tilde{Z}^{\prime}$ for the $U(1)_\chi$ model. In a scenario where  $\tilde{Z}^{\prime}$
is the next LSP (NLSP), eventually total strong production will become the cross-section of
$\tilde{Z}^{\prime}$ pair. The $\tilde{Z}^{\prime}$ then decays to a RHN  along with a LSP. 
In the next sections we shall be considering the recent experimental bounds to see the prospect
of these channels for further collider studies.

\section{Experimental bounds and Benchmark points}

The discovery of the Higgs boson with a mass around 125.5 GeV  \cite{Higgsd1,Higgsd2} has been very crucial in  understanding  the electroweak (EW) symmetry breaking. Although rather a large degree of fine-tuning cannot be avoided, SUSY still remains as a promising theory beyond SM stabilizing such a light Higgs boson mass. 
Various SUSY scenarios  need large quantum corrections to have the Higgs boson mass around 125.5 GeV, which gives strong bounds to the SUSY  mass spectrum contributing to the one-loop Higgs boson mass , $m_h$.  So the strongly interacting SUSY particles also get indirect bounds from the Higgs boson mass.  
It has been shown that for pMSSM one needs either very large stop masses or large splitting between the two mass eigen states in order to have $\sim 125$ GeV Higgs. \cite{carena} . In cMSSM/mSUGRA one needs squarks masses greater than few TeV \cite{cmssm}.  In this study we choose our parameter space to accommodate the lightest  Higgs boson mass around 125.5 GeV considering the theoretical uncertainties\footnote{Suspect \cite{suspect} and FeynHiggs \cite{feynhiggs} reportedly have around 3 GeV of uncertainties for low $\tan{\beta}\sim 5$. This is due to the uncertainty in the two-loop Higgs mass calculation and also reported by \cite{naturalsusy}}.

For the collider study we consider that recent bounds on third generation squark masses \cite{thrdgenb}. 
Most of the above bounds considers $\tilde{t_1} \to t \tilde{\chi^0_1}$,  $\tilde{b_1} \to b \tilde{\chi^0_1}$ or/and $\tilde{t_1} \to b\tilde{\chi^{+}_1}$, $\tilde{b_1} \to t \tilde{\chi^{-}_1}$ branching fractions to be unity.  So for a very light stop, one needs to go for a heavy LSP for standard decays as above.  The non-standard decays, such as $\tilde{t}_{1,2}(\tilde{b}_{1,2}) \to t (b) \tilde{Z'}$, where $\tilde{Z'}$ is not the LSP, will reduce the lower bounds further and reopen much lighter stop and sbottom masses. For our study we take relatively larger stop and sbottom masses given in Table~\ref{massbp}. We  also choose $m_{\tilde{g}}\geq 1.4$ TeV to satisfy recent gluino mass bound \cite{glnbd, djlhc}. The first two generations squarks masses we have taken more than a TeV \cite{djlhc}.

Table~\ref{bps} presents the input parameters chosen for the benchmark points for the collider study. The heavy  pseudo-scale boson mass $m_A$ is chosen to be 1 TeV, and thus all the heavy  Higgs bosons are decoupled 
from the analysis.


\begin{tiny}
\begin{table}[h]
\begin{center}
\renewcommand{\arraystretch}{1.0}
\hspace*{-1cm}
\begin{tabular}{||c||c|c|c|c|c|c|c|c|c|c|c|c||}
\hline
 &$M_{\tilde q_{1,2}}$& $M_{\tilde{Q}_3}$&$M_{\tilde{t}_R}$&$M_{\tilde{b}_R}$&
$\tan\!{\beta}$&$\mu$&$M_1$&$M_2$&$M_3$&$m_{\tilde{Z}^{\prime}}$&$A_t$&$A_b$\\
\hline
\hline
BP1 &1000&700&800&650&20&-730&700&750&1400&300&1600&1500\\
\hline
BP2 &1000&700&800&650&15&-130&230&400&1400&220&1625&1500\\
\hline
BP3 &2000&800&800&700&20&-730&700&750&2000&300&1600&1500\\
\hline
\hline
\end{tabular}
\caption{Input parameters (masses in GeV) for the benchmark points.}\label{bps}
\end{center}
\end{table}
\end{tiny}


Table~\ref{massbp}
shows the respective SUSY particle mass spectrum generated by Suspect \cite{suspect}. 
As we implement our vertices in CalcHEP \cite{calchep} which uses Suspect for SUSY spectrum generation.
One can see that for BP1 and BP3 $\tilde{Z'}$ is NLSP, whereas for BP2 it is next to next LSP (NNLSP). In BP3
first two generations of squarks and gluino are decoupled having masses $\sim 2$ TeV.  In all three benchmark point we consider a right-handed sneutrino $\tilde N_1$ as the LSP with $m_{\tilde N_1} = 110$ GeV.
We will take the corresponding right-handed neutrino mass to be $m_N=100$ GeV.


\begin{table}[h]
\begin{center}
\renewcommand{\arraystretch}{1.2}
\begin{tabular}{||c|c|c|c|c|c|c|c|c||}
\hline
& $m_{\tilde{t}_1}$&$m_{\tilde{t}_2}$&$m_{\tilde{b}_1}$&$m_{\tilde{b}_2}$
&$m_{\tilde{g}}$&$m_{\tilde{\chi}^0_1}$&$m_{\tilde{\chi}^\pm_{1}}$&$m_{\tilde Z'}$\\
\hline
\hline
BP1 &561.5&910.0&634.3&730.8&1400&671.1&693.1&300.0\\
\hline
BP2 &547.4&904.0&656.2&711.3&1400&117.3&129.5&220.0\\
\hline
BP3 &543.5&900.0&622.7&760.9&2000&677.2&699.6&300.0\\
\hline
\hline
\end{tabular}
\caption{Mass spectra (in GeV) for the benchmark points 
}\label{massbp}
\end{center}
\end{table}


\section{Production rates and decays}

We calculate the production rates of the quark and gluino pairs via CalcHEP \cite{calchep} at the LHC with the center of mass energy of 14 TeV. The renormalization and factorization scales are chose as
$m_{\tilde{t}_1}$ and CTEQ6L \cite{pdf} is chosen as parton distribution function (PDF). 
One can see from Table~\ref{crossbp}
that only lighter third generation squarks have relatively large cross-sections. We do not show the
cross-sections of first two generation squark pairs which are less than 10 fb.  


\begin{table}[h]
\begin{center}
\renewcommand{\arraystretch}{1.2}
\begin{tabular}{||c|c|c|c|c|c||}
\hline
 &$\tilde{t}_1\tilde{t}_1$&$\tilde{t}_2\tilde{t}_2$&$\tilde{b}_1\tilde{b}_1$&$\tilde{b}_2\tilde{b}_2$&$\tilde{g}\tilde{g}$  \\

\hline
BP1&176.29&10.0&88.63&38.70&3.68\\

\hline
BP2&200.00&10.05&82.66&45.86&3.68\\

\hline
BP3&213.6&6.19&65.78&19.03&1.52\\
\hline

\end{tabular}
\caption{Cross-sections at LHC14 in fb. }\label{crossbp}
\end{center}
\end{table}

Let us look at the $\tilde{Z}^{\prime}$ productions rates from the third generation SUSY cascade decay. Table~\ref{zpbr} gives the decay branching fractions of the different squarks
to $\tilde{Z}^{\prime}$. As explained in our earlier work that for
$U(1)_{\chi}$ model right handed squark will have larger decay
branching fraction to $\tilde{Z}^{\prime}$ as compared to the left
handed squarks \cite{mdl}.  In spite of having a larger enough
branching fraction $35 \sim 68\%$ from right-handed squarks, the first
two generations fail to contribute due to lager allowed
masses. Whereas for third generations mixing between the left handed
and right handed squakrs plays a role in reducing the effective
branching fraction to $\tilde{Z}^{\prime}$ substantially. For BP1 and
BP3 where $\tilde{Z}^{\prime}$ is the NLSP, Br($\tilde{t}_1\to t
\tilde{Z'}$) and  Br($\tilde{b}_1\to b \tilde{Z'}$) are
100\%. \footnote{For BP1 and BP3, where the $\tilde{Z'}$ is NLSP and
  there are also the secondary decays like $\tilde{t}_2\to h/Z, \tilde{t}_1$ and
  $\tilde{b}_2\to \tilde{t}_1 W$ which further contribute to
  $\tilde{Z'}$ with more jets and leptons. These secondary
  contributions will enhance the signal significance much higher.}

\begin{table}[h]
\begin{center}
\renewcommand{\arraystretch}{1.2}
\begin{tabular}{||c|c|c|c|c|c|c|c|c||}
\hline
 &$\tilde{u}_L$& $\tilde{u}_R$ & $\tilde{d}_L$& $\tilde{d}_R$ & $\tilde{t}_1$&$\tilde{t}_2$&$\tilde{b}_1$&$\tilde{b}_2$ \\

\hline
BP1&0.117&0.347&0.120&0.679&1.00&0.009&1.00&0.034\\
\hline
BP2&0.04&0.15&0.04&0.42&0.017&0.005&0.11&0.01\\
\hline

BP3&0.04&0.18&0.04&0.47&1.00&0.01&1.00&0.018\\
\hline

\end{tabular}
\caption{Decay branching fraction of squarks to ${\tilde{Z}^{\prime}}$}\label{zpbr}
\end{center}
\end{table}


The $\tilde{Z}^{\prime}$ thus produced will decay  via $\tilde{Z}^{\prime}\to N \tilde{N}_1$. 
The right-handed neutrino then decays through lepton
flavour violating $eW$, $Z\nu$ and $h \nu$ depending on $m_N$:
\bea\label{Ndc}
N &\to& e^\pm W^\mp  \quad (79\%) \\ \nonumber
&\to& \nu_e  Z  ~~~~\quad (21\%) \\ \nonumber
&\to& \nu_e h  ~~~~~ \quad (0\%)
\eea
where in the pararenthes are shown the branching fractions for $m_N=100$ GeV.
 For this study we only focus on the mode $N\to e^\pm W^\mp$ which produces same number of positively and negatively charged  electrons. 
 Thus from $pp\to NN +X$, we expect have charge symmetry considering only this lepton flavour violating decay.  

In  BP1 and BP3 as mentioned earlier $\tilde{t}_1$ and $\tilde{b}_1$ completely decays to
$\tilde{Z}'$ which further decays to  $N \tilde{N}_1$, and then
the right-handed neutrino prefers to decay into electron and $W^\pm$ boson.
The final state coming from the sbottom production and decay will have two $b$-jet and 4 non-$b$-jet at the partonic level if we demand both the $W$s to decay hadronically:
 \bea\label{fs1}
\tilde{b}_{1,2}&\to& b \tilde{Z}^{\prime}\to b N \tilde{N}_1\to b e W \ptmiss\\\nonumber
\tilde{b}_{1,2}\tilde{b}^*_{1,2}&\to& 2 e + 2 b +4q + \ptmiss .
\eea
Similarly, for $\tilde{t}_{1,2}$ we have
\bea\label{fs4}
\tilde{t}_{1,2}&\to& t \tilde{Z}'\to bW N\ntrl1\to b +2W + e + \ptmiss \\ \nonumber
\tilde{t}_{1,2}\tilde{t}^*_{1,2}&\to&  2e +2b + 8q + \ptmiss .
\eea
Thus, it will be our primary interest to look for the lepton flavour violating final state:
\be\label{fs2}
 (2e-2\mu) + 2 b +n_{q} + \ptmiss.
\ee
with $n_{q} \geq 4$ for both same-sign or opposite-sign $e$ or $\mu$.
In addition, 3$l$ and 4$l$ signatures coming from leptonic decays of $W$ are also promising to look for the signal events. Here, let us note that the similar di-electron signals, but with smaller number of jets, can appear also in the model of Allanach et.al.~\cite{allanach}. Additional lepton/jet and missing energy signatures could be useful features distinquishing different models.

Unlike BP1 and BP3, in BP2  $\tilde{Z'}$ is not NLSP but NNLSP (see Table~\ref{massbp}) and the NLSP
 is of  the Higgsino type. This results in sharing the third generation squark branching with the higgsino-like lighter neutralino ($\tilde{\chi}^0_1$) and lighter chargino ($\tilde{\chi}^\pm_1$). The effect can be seen from Table~\ref{zpbr}, which makes BP2 more challenging.

\section{LHC Phenomenology}

As discussed, relatively light third generation sqaurks can give rise to to numerous 
flavour violating dilepton final states in association with some $b$-jets and non-$b$-jets mainly coming from $W$ bosons along with the missing energy.  It is also important to look for the Majorana nature of the RHN which decays to both sign of electrons, i.e., $N\to e^\pm W^\mp $. This suggests that determining the charge multiplicity we should expect to  have similar number of lepton flavour violating events for both OSD and SSD. 

When some of the $W$s decay leotonically these give rise to $3\ell$ and $4\ell$ signatures. In this collider study we mainly focus on the dilepton, trilepton and $4\ell$ final states. For this purpose
we generated the events in CalcHEP \cite{calchep} and simulated with {\tt PYTHIA} \cite{pythia} via the the SLHA interface \cite{slha} for the decay branching and mass spectrum. 

For hadronic level simulation we have used {\tt Fastjet-3.0.3} \cite{fastjet} algorithm for the jet formation with the following criteria:
\begin{itemize}
  \item the calorimeter coverage is $\rm |\eta| < 4.5$

  \item $ p_{T,min}^{jet} = 20$ GeV and jets are ordered in $p_{T}$
  \item leptons ($\rm \ell=e,~\mu$) are selected with
        $p_T \ge 10$ GeV and $\rm |\eta| \le 2.5$
  \item no jet should match with a hard lepton in the event
   \item $\Delta R_{lj}\geq 0.4$ and $\Delta R_{ll}\geq 0.2$
  \item Since efficient identification of the leptons is crucial for our study, we additionally require  
hadronic activity within a cone of $\Delta R = 0.3$ between two isolated leptons to be $\leq 0.15 p^{\ell}_T$ GeV in the specified cone.

\end{itemize}

\begin{figure}[hbt]
\begin{center}

\includegraphics[width=0.495\linewidth]{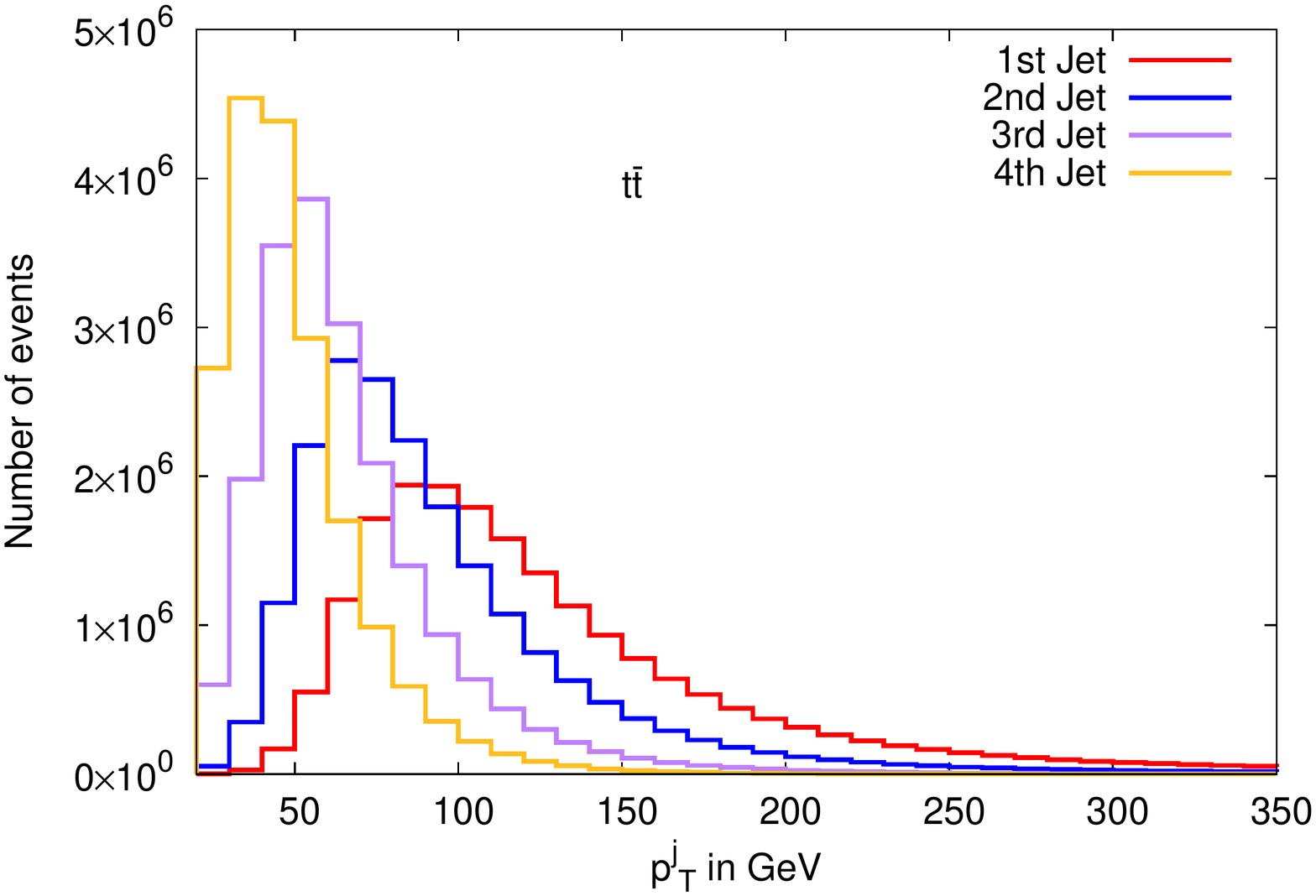}
\includegraphics[width=0.495\linewidth]{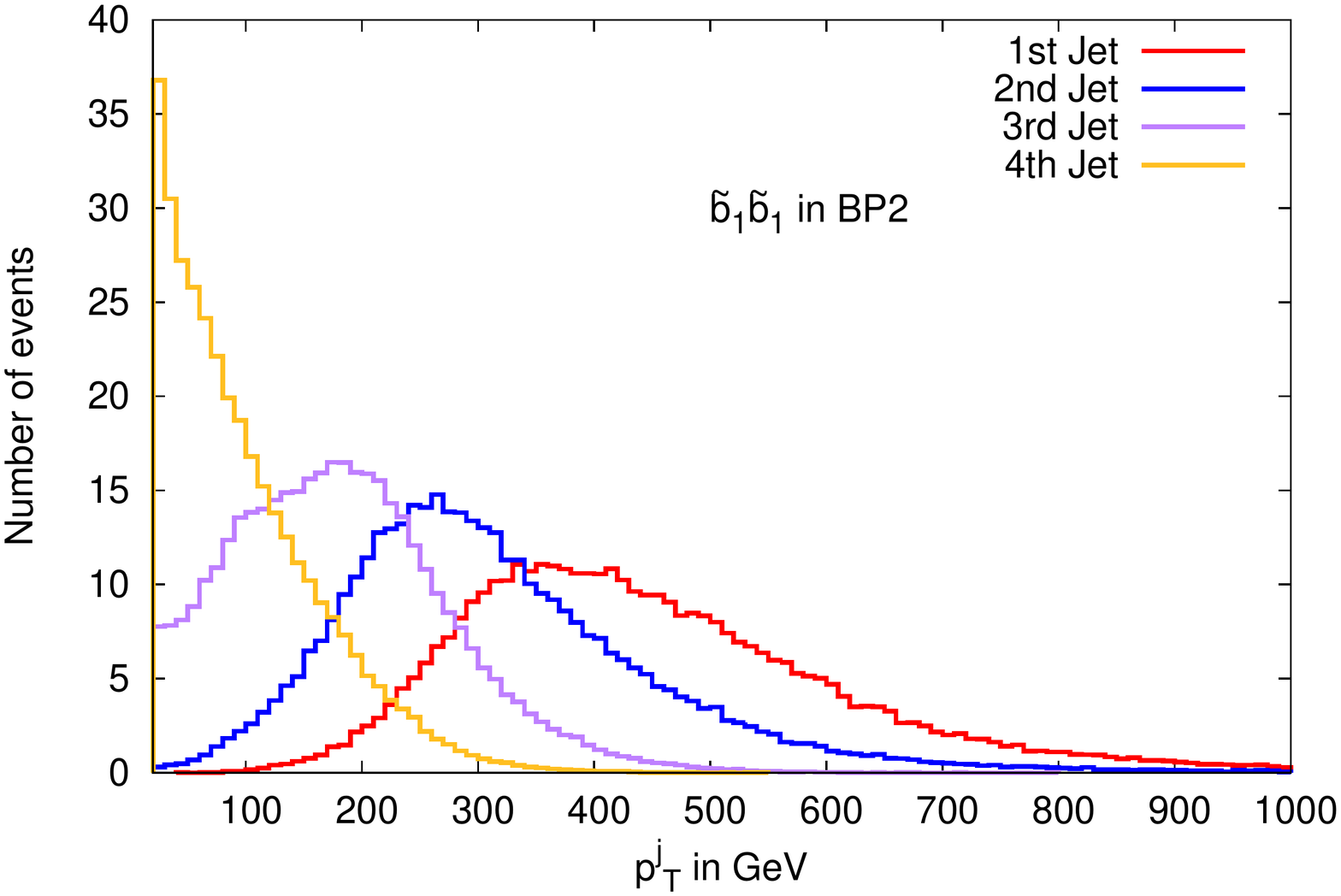}
\caption{Ordered $p^{jet}_T$ distribution for $t\bar{t}$ (left) and for $\tilde{b}_1\tilde{b}^*_1$ in BP2  (right) at an integrated luminosity of 50 fb$^{-1}$.}\label{ptj}

\end{center}
\end{figure}

We show in Figure~\ref{ptj} the jet $p_T$ distributions coming from  $t\bar{t}$ (left) and $\tilde{b}_1\tilde{b}^*_1$ (right) for BP2.  It is clear that the jets coming from $\tilde{b}_{1}$ decay could be as hard as $p_T\gsim 300$ GeV, which is very unlikely in the case of $t\bar{t}$. 

\begin{figure}[hbt]
\begin{center}

\includegraphics[width=0.495\linewidth]{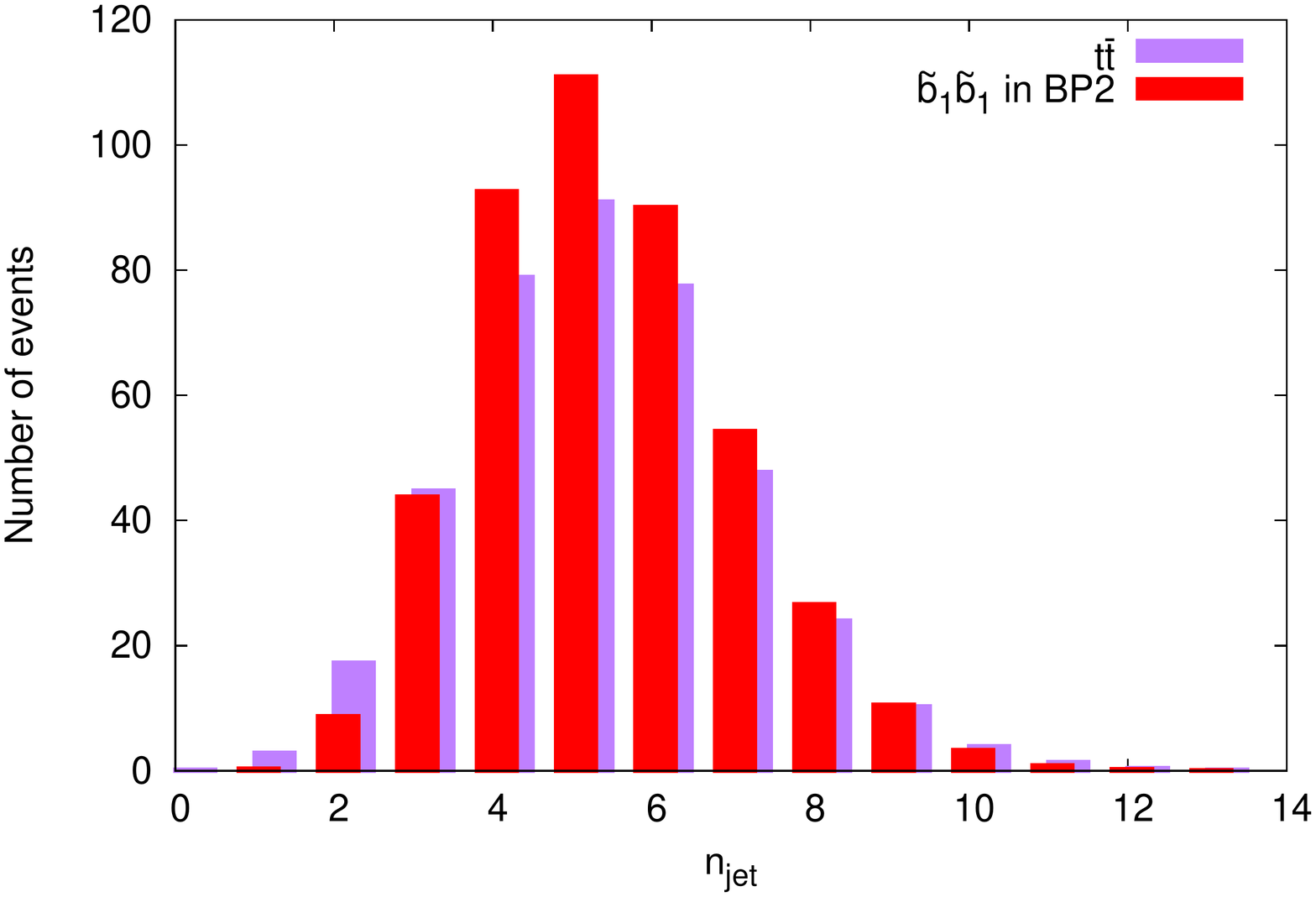}
\includegraphics[width=0.495\linewidth]{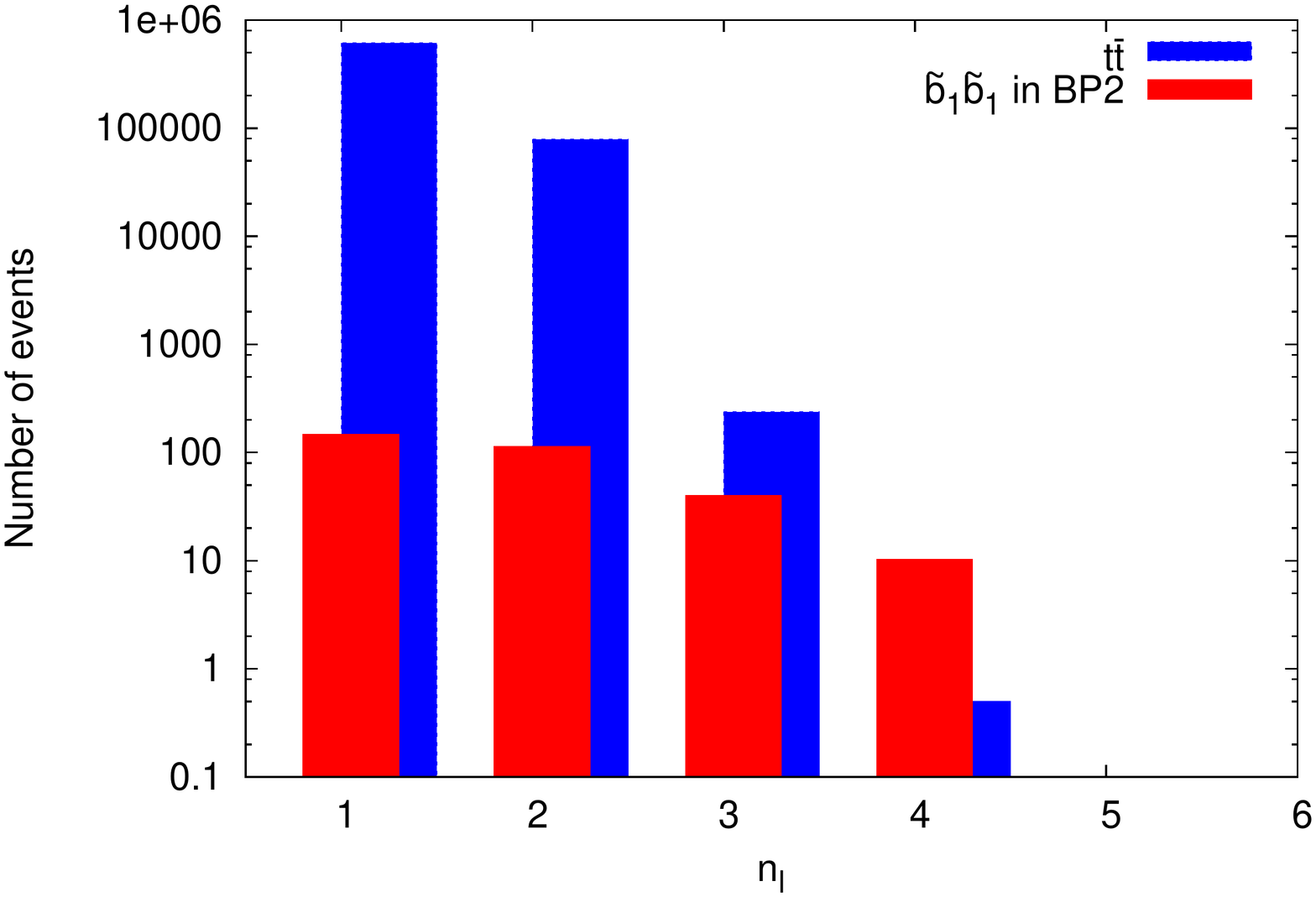}
\caption{ $n_{\rm{jet}}$ distribution (left) and $n_{\ell}$ distribution (right) for $t\bar{t}$ and for $\tilde{b}_1\tilde{b}^*_1$ in BP2 at an integrated luminosity of 50 fb$^{-1}$.}\label{njl}

\end{center}
\end{figure}

Figure~\ref{njl} shows the jet (left) and lepton (right) multiplicity distribution for  $\tilde{b}_1\tilde{b}^*_1$ (BP2) and for the dominant background $t\bar{t}$. We can see that though both the signal and backgrounds can have large number of jets in the final states, the  $\tilde{b}_1\tilde{b}^*_1$ can have relatively large numbers of $3\ell$ and $4\ell$ final states. 

\begin{figure}[hbt]
\begin{center}

\includegraphics[width=0.495\linewidth]{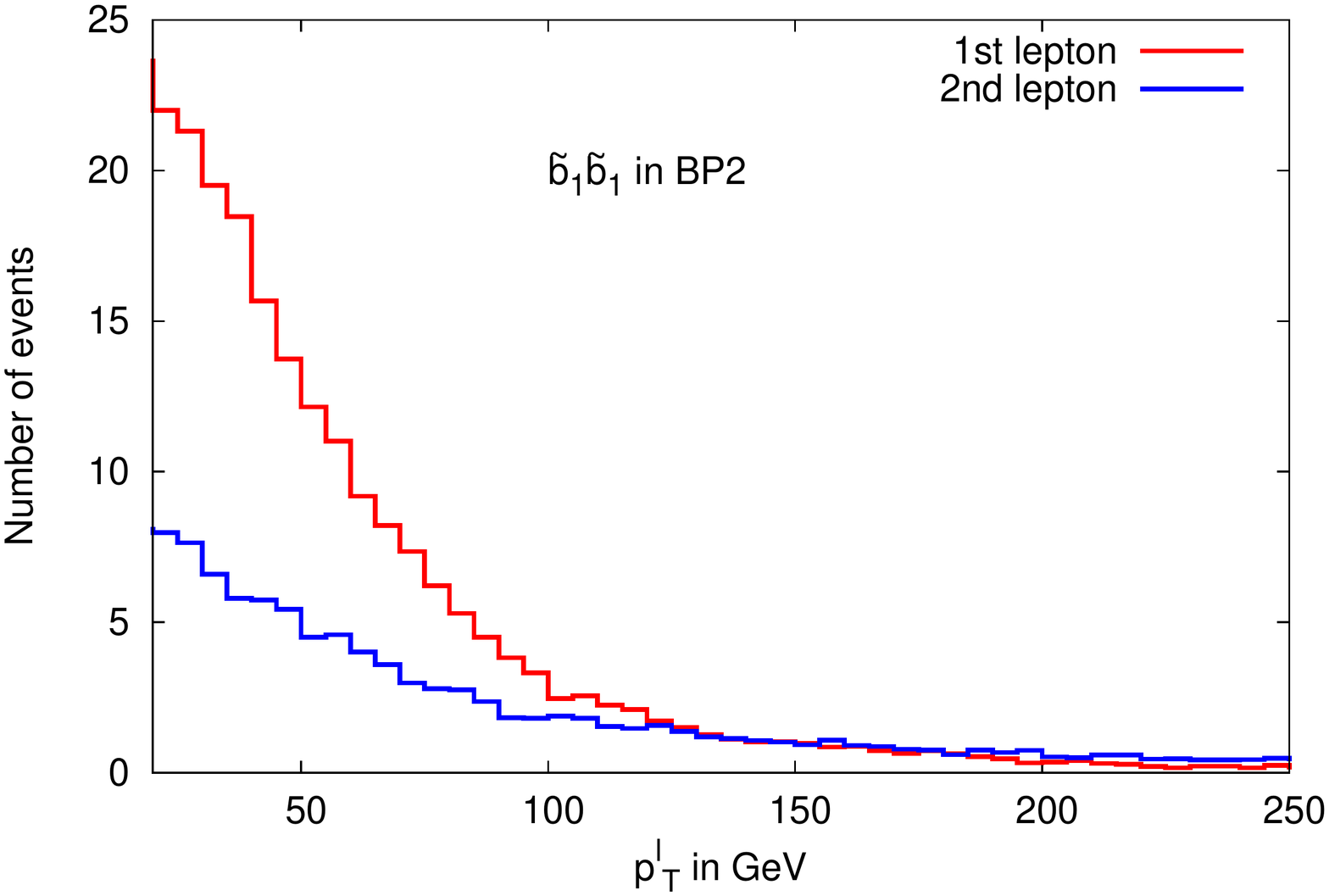}
\includegraphics[width=0.495\linewidth]{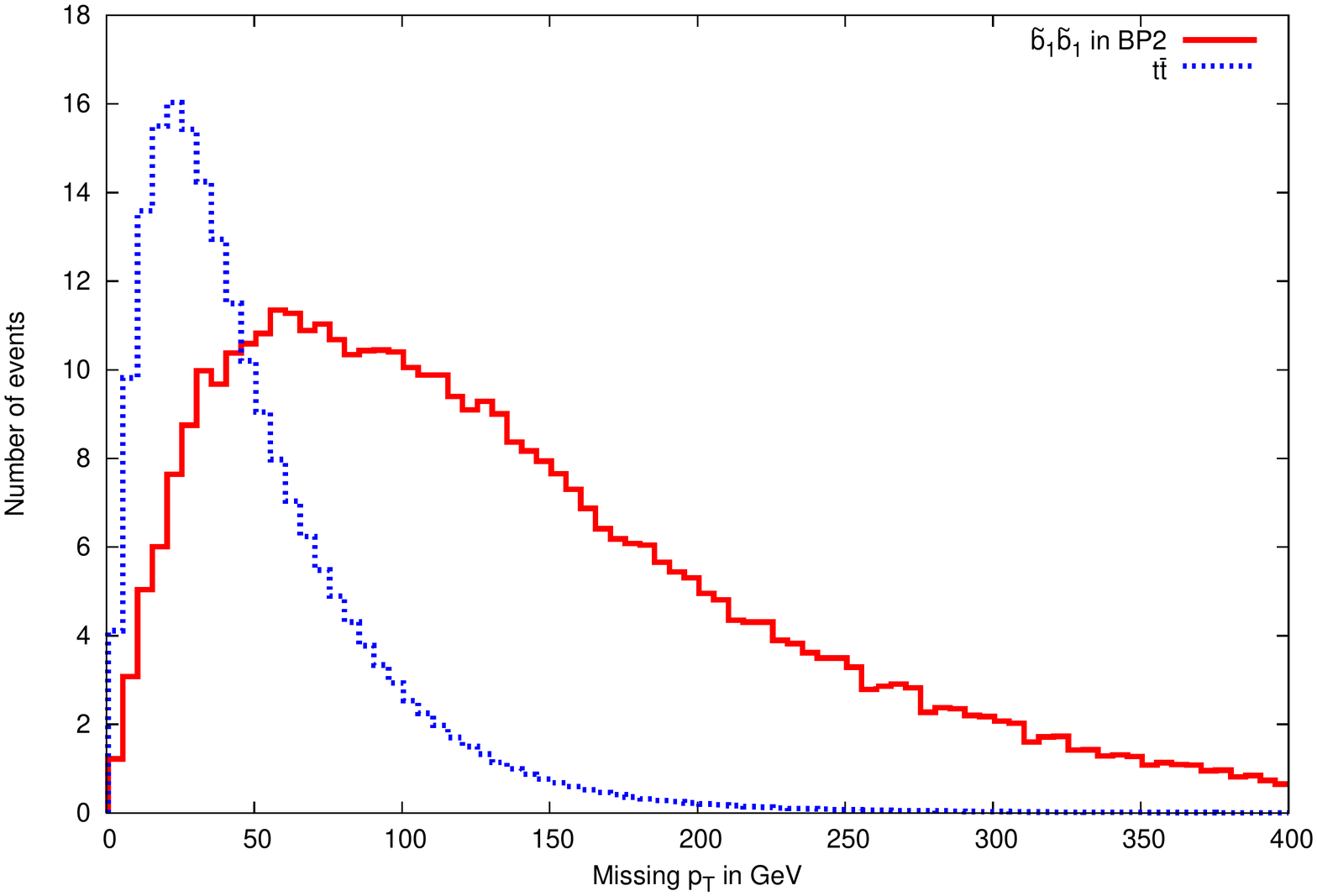}
\caption{Lepton $p_T$ distribution for BP2 $\tilde{b}_1\tilde{b}^*_1$ (left), and $\ptmiss$ distribution for $t\bar{t}$ and for BP2 $\tilde{b}_1\tilde{b}^*_1$ (right) at an integrated luminosity of 50 fb$^{-1}$.}\label{mispt}

\end{center}
\end{figure}

From Figure~\ref{mispt} (left) we can see that the leptons coming from 
$\tilde{b}_1\tilde{b}^*_1$ have a high energy tail. The contribution of this 
high energy tail is coming from the decay the right-handed neutrino to $eW$, which gets
the boost from the cascade decays of $\tilde{b}_1$.
 Figure~\ref{mispt} (right) presents the $\ptmiss$ distributions
for the signal $\tilde{b}_1\tilde{b}^*_1$ and for the dominant background $t\bar{t}$. It is clear the LSP in the case of  $\tilde{b}_1\tilde{b}^*_1$, adds to large missing $p_T$ as compared to the neutrinos for $t\bar{t}$.  A $\ptmiss$ cut of $\ptmiss \gsim 100$ GeV will kill most of the SM backgrounds.

In this article we focus on the multilepton final states with lepton number and flavour violation. In the following subsections we describe the final states with their signal and backgrounds number at LHC14.

\subsection{$2e$ and $2\mu$ signatures and charge multiplicity}

Our signal signature shown in  Eq.~\ref{fs1} involves the right handed neutrino decay to $eW(jj)$
leading to $2e+2b+4j+\ptmiss$ in the final state.  In the process of hadronization and jet formation with ISR/FSR, more number of jets are produced.  We have also seen that the right-handed neutrino can decay either $e^+ W^-$ or $e^- W^+$, $pp\to\tilde{b}_1\tilde{b}_1^* $ should generate both same sign and opposite sign di-lepton signatures. To extract out the lepton flavour asymmetry of electron and muon we look for final states where the $W^\pm$s decay hadronically. This implies to look for a final states $2e/2\mu +n_j$$\geq$6  
($n_b$$\geq$2).

 In Table~\ref{2elca} we show the number of events for $2e + n_j$$\geq$6  ($n_b$$\geq$2) final states for the signal benchmark points and the SM backgrounds at 50 fb$^{-1}$ of luminosity of 14 TeV LHC. We consider  $t\bar{t}Z$,  $t\bar{t}W$ and $t\bar{t}b\bar{b}$  as the dominant SM backgrounds which are not reducible backgrounds with ISR/FSR. We show the the contributions coming from each third generation sqaurks pair production process for both SSD and OSD, and also from the SM backgrounds. The respective signal significance are also been calculated and listed.  From Table~\ref{2elca} we can clearly see that the signal numbers are symmetric in SSD and OSD, whereas the backgrounds prefer OSD as expected.  The  backgrounds come  from the two opposite sign $W$ decay or from one neutral gauge boson ($Z$) decay. The slightly large number of OSDs in the case of signal happens due to the decay kinematics of the right-handed neutrinos.  Generally the charge symmetry is maintained when we tag two leptons coming from two different right-handed neutrinos. When $N$ decays to $e^\pm W^\mp$, sometimes one of the electrons can not be isolated from the jets coming from the associated $W$ boson, i.e., cannot  pass the jet-lepton isolation criterion ($\Delta R_{lj}> 0.4$ ). The original $3e$ events become
 $2e$ events where the second right-handed neutrino decay ($N \to e^\pm W^\mp$) can contribute to dilepton final states with the leptonic decay of the associated $W$ boson. This would always be of opposite-sign leptons as the right-handed neutrino ($N$) is charge neutral.  Thus single right-handed neutrino contributing to dilepton final sates makes it opposite-sign. Relaxing the isolation criterion reduces the discrepancy.  
 
As expected, BP1 and BP3 produce much more signal numbers than BP2 because both $\tilde{t}_1$ and $\tilde{b}_1$ fully decays to $\tilde{Z'}$.  In the backgrounds $t\bar{t}$ makes highest contribution due to its large cross-section. For opposite-sign di-electron final states BP1 and BP3  have more than $8 \sigma $ significance, whereas for the same-sign di-electron  they are around 
$21 \sigma$. BP2
 fails to cross even $2\sigma$.


\begin{table}
\begin{center}
\renewcommand{\arraystretch}{1.2}
\begin{tabular}{||c||c|c|c|c||c|c|c|c||}

\hline\hline 14TeV/50fb$^{-1}$ &\multicolumn{4}{|c|}{Signal}&\multicolumn{4}{|c|}{Background}\\\hline
$2e +n_{j} \geq 6$ &Charge &BP1&BP2 &BP3 &$t\bar{t}Z$& $t\bar{t}W$ & $t\bar{t}b\bar{b}$& $t\bar{t}$ \\
$(n_b \geq 2) $&Multiplicity &&&&&&& \\
\hline \hline \multirow{2}{*}{$\tilde{t}_1\tilde{t}^*_1$}&OSD &619.45 &12.88 &731.24&\multirow{5}{*}{113.18}&\multirow{5}{*}{15.53}&\multirow{5}{*}{0.87}&\multirow{5}{*}{9220.04}\\
&SSD&457.12 &9.19 &522.48&\multirow{5}{*}{3.76}&\multirow{5}{*}{9.12}&\multirow{5}{*}{0.87}&\multirow{5}{*}{236.54}\\
 \multirow{2}{*}{$\tilde{b}_1\tilde{b}^*_1$}&OSD  &206.69 &29.85&158.52& &&& \\
&SSD  &153.27 &21.12&115.34& &&& \\
 \multirow{2}{*}{$\tilde{b}_2\tilde{b}^*_2$}&OSD  &3.51 &1.37 &0.89&&&& \\
&SSD  &2.44 &0.87 &0.63&&&& \\
\multirow{2}{*}{ $\tilde{t}_2\tilde{t}^*_2$}&OSD  &0.34&0.22 &0.25 &&&& \\ 
&SSD  &0.24&0.14 &0.17 &&&& \\ 
\hline
\multirow{2}{*}{Total} &OSD&830.00&44.32&890.90&\multicolumn{4}{|c|}{9349.62}\\
&SSD&613.07&31.32&638.62&\multicolumn{4}{|c|}{250.29}\\
\hline
\multirow{2}{*}{Significance} & OSD &8.22 &0.46&8.80&\multicolumn{4}{|c|}{} \\
 &  SSD&20.86 &1.87&21.42&\multicolumn{4}{|c|}{} \\

\hline
\hline
\end{tabular}
\caption{Number of events in $2e +n_{j}$$\geq$6 ($n_b$$\geq$2)  final states for  the benchmark points and the SM backgrounds at LHC14 with an integrated luminosity of 50 fb$^{-1}$.}
\label{2elca}
\end{center}
\end{table}

Table~\ref{2mlca} presents corresponding number of di-muon events for the benchmark points and the backgrounds.  As the right-handed neutrino decays only to electron flavour, the number of events 
for the muon final states are very low. The backgrounds numbers are similar to the electron final states in Table~\ref{2elca}.


\begin{table}
\begin{center}
\renewcommand{\arraystretch}{1.2}
\begin{tabular}{||c||c|c|c|c||c|c|c|c||}

\hline\hline 14TeV/50fb$^{-1}$ &\multicolumn{4}{|c|}{Signal}&\multicolumn{4}{|c|}{Background}\\\hline
$2\mu +n_{j} \geq 6$ &Charge &BP1&BP2 &BP3 &$t\bar{t}Z$& $t\bar{t}W$ & $t\bar{t}b\bar{b}$& $t\bar{t}$ \\
$(n_b \geq 2) $&Multiplicity &&&&&&& \\
\hline \hline \multirow{2}{*}{$\tilde{t}_1\tilde{t}^*_1$}&OSD &27.67 &0.58 &35.45&\multirow{5}{*}{120.70}&\multirow{5}{*}{16.77}&\multirow{5}{*}{1.46}&\multirow{5}{*}{9121.49}\\
&SSD&7.40 &0.14 &8.78&\multirow{5}{*}{4.42}&\multirow{5}{*}{7.56}&\multirow{5}{*}{1.46}&\multirow{5}{*}{251.32}\\
 \multirow{2}{*}{$\tilde{b}_1\tilde{b}^*_1$}&OSD  &4.79 &1.29&3.41& &&& \\
&SSD  & 0.20&0.33&0.12& &&& \\
 \multirow{2}{*}{$\tilde{b}_2\tilde{b}^*_2$}&OSD  &0.09 &0.03 &0.02&&&& \\
&SSD  &0.003 &0.00 &0.001&&&& \\
\multirow{2}{*}{ $\tilde{t}_2\tilde{t}^*_2$}&OSD  &0.02&0.01 & 0.01&&&& \\ 
&SSD  &0.005&0.003 &0.003 &&&& \\ 
\hline
\multirow{2}{*}{Total} &OSD&117.28&7.01&106.36&\multicolumn{4}{|c|}{9260.42}\\
&SSD&7.0&0.46&8.91&\multicolumn{4}{|c|}{264.76}\\
\hline
\multirow{2}{*}{Significance} & OSD &1.21&0.07&1.10&\multicolumn{4}{|c|}{} \\
 &  SSD&0.42 &0.03&0.54&\multicolumn{4}{|c|}{} \\

\hline
\hline
\end{tabular}
\caption{Number of events for $2\mu +n_{j}$$\geq$6 ($n_b$$\geq$2)  final states for  the benchmark points and the SM backgrounds at LHC14 with an integrated luminosity of 50 fb$^{-1}$.
TeV.}\label{2mlca}
\end{center}
\end{table}

Next we take the difference in number of events between electrons and muons for both OSD and SSD. Table~\ref{2e2mlca} shows the event numbers in the $(2e-2\mu) +n_{j}$$\geq$5 ($n_b$$\geq$2)  final state for all the benchmark points and the backgrounds. For BP1 and BP3 we can have around $25 \sigma$ signal significance for OSD and SSD flavour difference. It is encouraging to see that the significance of BP3 can reach to about $7\sigma$ for SSD.


\begin{table}
\begin{center}
\renewcommand{\arraystretch}{1.2}
\begin{tabular}{||c||c|c|c||c|c|c|c||}
\hline\hline 14TeV/50fb$^{-1}$ &\multicolumn{3}{|c|}{Signal}&\multicolumn{4}{|c|}{Background}\\\hline
$(2e-2\mu) +n_{j} \geq 6$ &BP1&BP2 &BP3 &$t\bar{t}Z$& $t\bar{t}W$ & $t\bar{t}b\bar{b}$& $t\bar{t}$ \\
\hline \hline 
OSD &712.72 &37.31 &784.54&-7.52&-1.24&-0.59&98.55\\
\hline
SSD&606.07 &30.86 &629.71&-0.66&1.56&-0.59&-14.78\\
\hline
\multirow{2}{*}
{Significance}{~~OSD}& 25.17&3.31&26.54&\multicolumn{4}{|c|}{} \\
~~~~~~~~~~~~~~~~ {SSD} & 24.91&7.62&25.39&\multicolumn{4}{|c|}{} \\
\hline
\hline
\end{tabular}
\caption{Number of events for $(2e-2\mu) +n_{jets} \geq 6(n_b\geq 2) $ final states for  the benchmark points and the SM backgrounds at LHC14 with an integrated luminosity of 50 fb$^{-1}$.}\label{2e2mlca}
\end{center}
\end{table}

\subsection{$3\ell$ signature}

Let us now consider the $3e$ final state, which is possible if one of the $W$s from the decay of the right-handed neutrino, decays leptonically. In this case we can have final state $3e +n_{j}$$\geq$4 ($n_b$$\geq$2)  from the decay of $\tilde{b}_1\tilde{b}_1^*$.  
Here, we also impose a missing energy cut  $\ptmiss \geq 100$ GeV to reduce the SM backgrounds. 

Table~\ref{sig3e} shows the number of events for this final state at an integrated luminosity of 50 fb$^{-1}$ for the benchmark points and the SM backgrounds.   BP1 and BP3 could reach for a signal significance of around 9 and 8$\sigma$ respectively. In case of BP2 due to the small branching fraction of $\tilde{b}_1 \to b \tilde{Z}'$ and  $\tilde{t}_1 \to t \tilde{Z}'$,  one needs larger luminosity to probe this $3e$ signal state.


\begin{table}
\begin{center}
\renewcommand{\arraystretch}{1.2}
\begin{tabular}{||c||c|c|c||c|c|c|c||}

\hline\hline 14TeV/50fb$^{-1}$ &
\multicolumn{3}{|c|}{Signal}&\multicolumn{4}{|c|}{Background}\\\hline
$3e +n_{j} \geq 4$ & BP1  & BP2 & BP3  &$t\bar{t}Z$&$t\bar{t}W$ &$t\bar{t}b\bar{b}$&$t\bar{t}$ \\
$(n_b \geq 2) + \ptmiss \geq 100$ GeV&   & &  &  & & & \\
\hline \hline \multirow{5}{*}{}$\tilde{t}_1\tilde{t}^*_1$ &38.86 &0.95 &37.03&\multirow{5}{*}{3.76}&\multirow{5}{*}{0.74}&\multirow{5}{*}{0.00}&\multirow{5}{*}{4.93}\\
 $\tilde{b}_1\tilde{b}^*_1$  &46.62 &3.50 &31.89 &&&& \\
 $\tilde{b}_2\tilde{b}^*_2$  & 0.86&0.35 &0.24  &&&& \\

 $\tilde{t}_2\tilde{t}^*_2$  &0.04&0.03 & 0.03&&&& \\ 
\hline
Total &86.39&4.84&69.20&\multicolumn{4}{|c|}{9.43}\\
\hline

Significance &8.83  & 1.28   & 7.80&\multicolumn{4}{|c|}{} \\

\hline \hline
\end{tabular}
\caption{Number of events for $3e +n_{j}$$\geq$4 ($n_b$$\geq$2) $+ \ptmiss \geq 100$ GeV final states for  the benchmark points and the SM backgrounds at LHC14 with an integrated luminosity of 50 fb$^{-1}$.}\label{sig3e}
\end{center}
\end{table}

Table~\ref{sig3mu} presents the number of events corresponding to the $3\mu$ final state. As expected, the signal could not contribute much for the $3\mu$ final state. Thus, the difference in the electron and muon events are of the same order of $3e$ final state as can be read from Table~\ref{sig3e3mu}. The signal significance gets slightly enhanced
as compared to $3e$ final states for all benchmark points and BP2 makes it to $3\sigma$.

\begin{table}
\begin{center}
\renewcommand{\arraystretch}{1.2}
\begin{tabular}{||c||c|c|c||c|c|c|c||}

\hline\hline 14TeV/50fb$^{-1}$ &
\multicolumn{3}{|c|}{Signal}&\multicolumn{4}{|c|}{Background}\\\hline
$3\mu +n_{j} \geq 4$ & BP1  & BP2 & BP3 &$t\bar{t}Z$&$t\bar{t}W$ &$t\bar{t}b\bar{b}$&$t\bar{t}$ \\
$(n_b \geq 2 ) + \ptmiss \geq 100$ GeV&  &  &  &  & & & \\
\hline \hline \multirow{5}{*}{}$\tilde{t}_1\tilde{t}^*_1$ &1.32 &0.03 &1.48&\multirow{5}{*}{6.00}&\multirow{5}{*}{0.74}&\multirow{5}{*}{0.00}&\multirow{5}{*}{4.93}\\
 $\tilde{b}_1\tilde{b}^*_1$  &0.36 &0.04 &0.25 &&&& \\
 $\tilde{b}_2\tilde{b}^*_2$  &0.005 &0.002 &0.001&&&& \\

 $\tilde{t}_2\tilde{t}^*_2$  &0.00&$5.6\times10^{-4}$&$5.9\times10^{-4}$   &&&& \\ 
\hline
Total &1.69&0.08&1.73&\multicolumn{4}{|c|}{11.67}\\
\hline

Significance & 0.46 & 0.02& 0.47&\multicolumn{4}{|c|}{} \\

\hline \hline
\end{tabular}

\caption{Number of events for $3\mu +n_{j}$$\geq$4 ($n_b$$\geq$2) $+ \ptmiss \geq 100$ GeV final states for  the benchmark points and the SM backgrounds at LHC14 with an integrated luminosity of 50 fb$^{-1}$.}\label{sig3mu}
\end{center}
\end{table}

\begin{table}
\begin{center}
\renewcommand{\arraystretch}{1.2}
\begin{tabular}{||c||c|c|c||c|c|c|c||}

\hline\hline 14TeV/50fb$^{-1}$ &
\multicolumn{3}{|c|}{Signal}&\multicolumn{4}{|c|}{Background}\\\hline
$(3e-3\mu) +n_{j} \geq 4$ & BP1  & BP2 & BP3  &$t\bar{t}Z$&$t\bar{t}W$ &$t\bar{t}b\bar{b}$&$t\bar{t}$ \\
$(n_b \geq 2) + \ptmiss \geq 100$ GeV&  & & &    & & & \\
\hline \hline \multirow{5}{*}{}$\tilde{t}_1\tilde{t}^*_1$ &37.54 &0.92 &35.55&\multirow{5}{*}{-2.24}&\multirow{5}{*}{0.00}&\multirow{5}{*}{0.00}&\multirow{5}{*}{0.00}\\
 $\tilde{b}_1\tilde{b}^*_1$  &46.26 &3.46 &31.64 &&&& \\
 $\tilde{b}_2\tilde{b}^*_2$  &0.86 &0.35 &0.24 &&&& \\

 $\tilde{t}_2\tilde{t}^*_2$  &0.04&0.03 & 0.04&&&& \\ 
\hline
Total &84.71&4.15&67.47&\multicolumn{4}{|c|}{-2.24}\\
\hline

Significance & 9.33& 3.00& 8.35  &\multicolumn{4}{|c|}{} \\

\hline \hline
\end{tabular}
\caption{Number of events for $(3e-3\mu) +n_{j}$$\geq$4 ($n_b$$\geq$2) $+ \ptmiss \geq 100$ GeV final states for  the benchmark points and the SM backgrounds at LHC14 with an integrated luminosity of 50 fb$^{-1}$.}\label{sig3e3mu}
\end{center}
\end{table}

\subsection{$4\ell$ signature}
Finally we are interested where two $W$s from the decays of the right-handed neutrinos (see Eq.~\ref{fs1}), decay leptonically. Here we do not distinguish the flavours of the charged lepton and consider both $e$ and $\mu$ in the final state. Of course, among these $4\ell$, two of them are electrons coming the flavour violating decays of the right-handed neutrino, i.e., $N\to e W$.  Table~\ref{sig4l} presents the number of events for the $4\ell +n_{j}$$\geq$3 ($n_b$$\geq$2) final state for the benchmark points and the SM backgrounds with an integrated luminosity of 50fb$^{-1}$ at the LHC14 . BP1 and BP3 reach more than $12\sigma$ significance but BP2 fails to contribute for this final state.

\begin{table}
\begin{center}
\renewcommand{\arraystretch}{1.2}
\begin{tabular}{||c||c|c|c||c|c|c|c||}

\hline\hline 14TeV/50fb$^{-1}$ &
\multicolumn{3}{|c|}{Signal}&\multicolumn{4}{|c|}{Background}\\\hline
$4\ell +n_{j} \geq 3$ & BP1  & BP2 & BP3 &$t\bar{t}Z$&$t\bar{t}W$ &$t\bar{t}b\bar{b}$&$t\bar{t}$ \\
$(n_b \geq 2) $&  & & &    & & & \\
\hline \hline \multirow{5}{*}{}$\tilde{t}_1\tilde{t}^*_1$ &122.15 &2.47 &143.26&\multirow{5}{*}{7.45}&\multirow{5}{*}{0.00}&\multirow{5}{*}{0.00}&\multirow{5}{*}{0.00}\\
 $\tilde{b}_1\tilde{b}^*_1$  & 35.14&5.92 &24.26 &&&& \\
 $\tilde{b}_2\tilde{b}^*_2$  &0.53 &0.20 &0.14 &&&& \\
 $\tilde{t}_2\tilde{t}^*_2$  &0.06&0.04 &0.05 &&&& \\ 
\hline
Total &157.88&8.64&167.71&\multicolumn{4}{|c|}{7.45}\\
\hline

Significance & 12.28& 0.25&12.67 &\multicolumn{4}{|c|}{} \\

\hline \hline
\end{tabular}
\caption{Number of events for $4\ell +n_j$$\geq$3 ($n_b$$\geq$2) final states for  the benchmark points and the SM backgrounds at LHC14 with an integrated
luminosity of 50 fb$^{-1}$.}\label{sig4l}
\end{center}
\end{table}

\subsection{Reconstruction of RHN}

\begin{figure}[hbt]
\begin{center}
\includegraphics[width=0.495\linewidth]{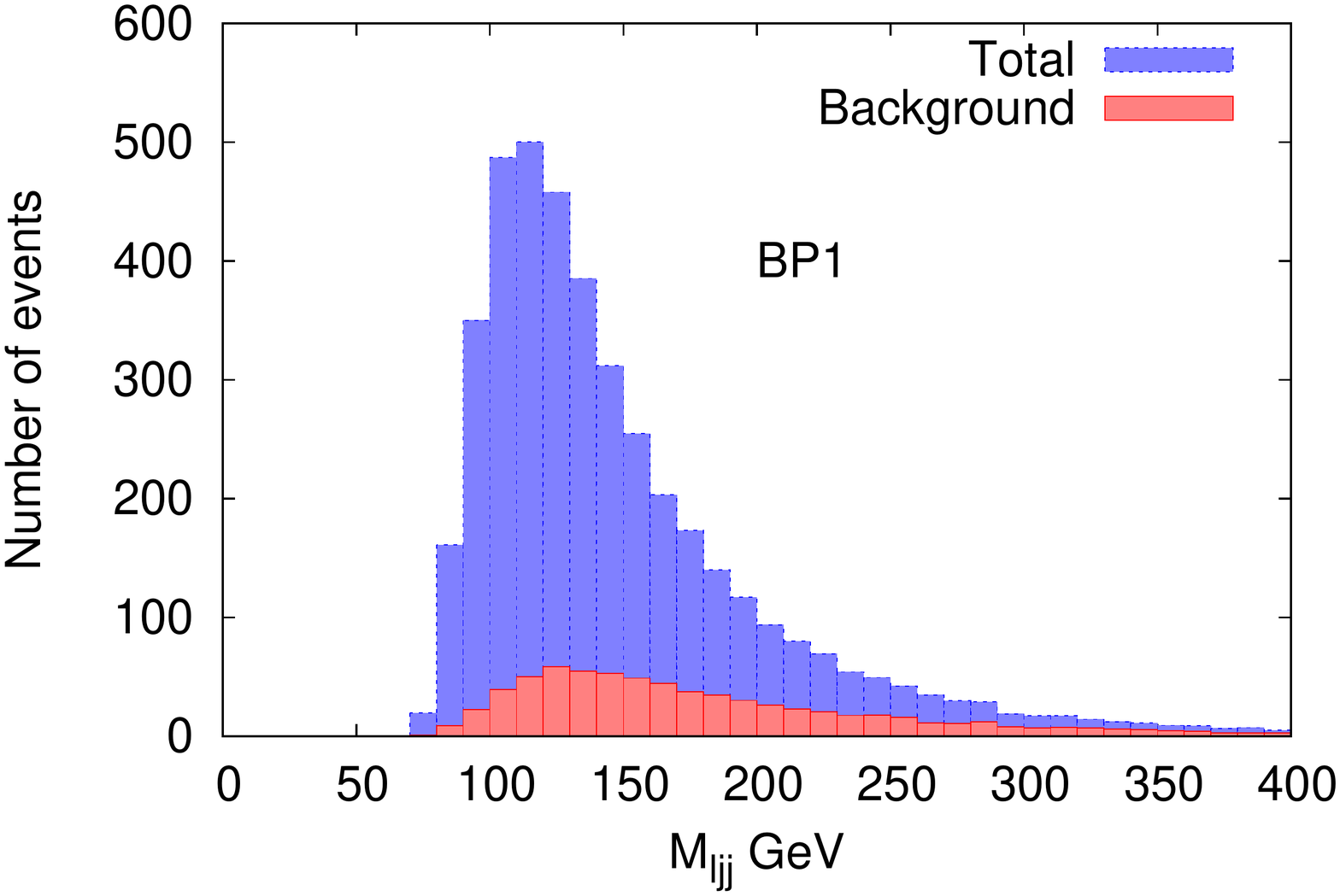}
\includegraphics[width=0.495\linewidth]{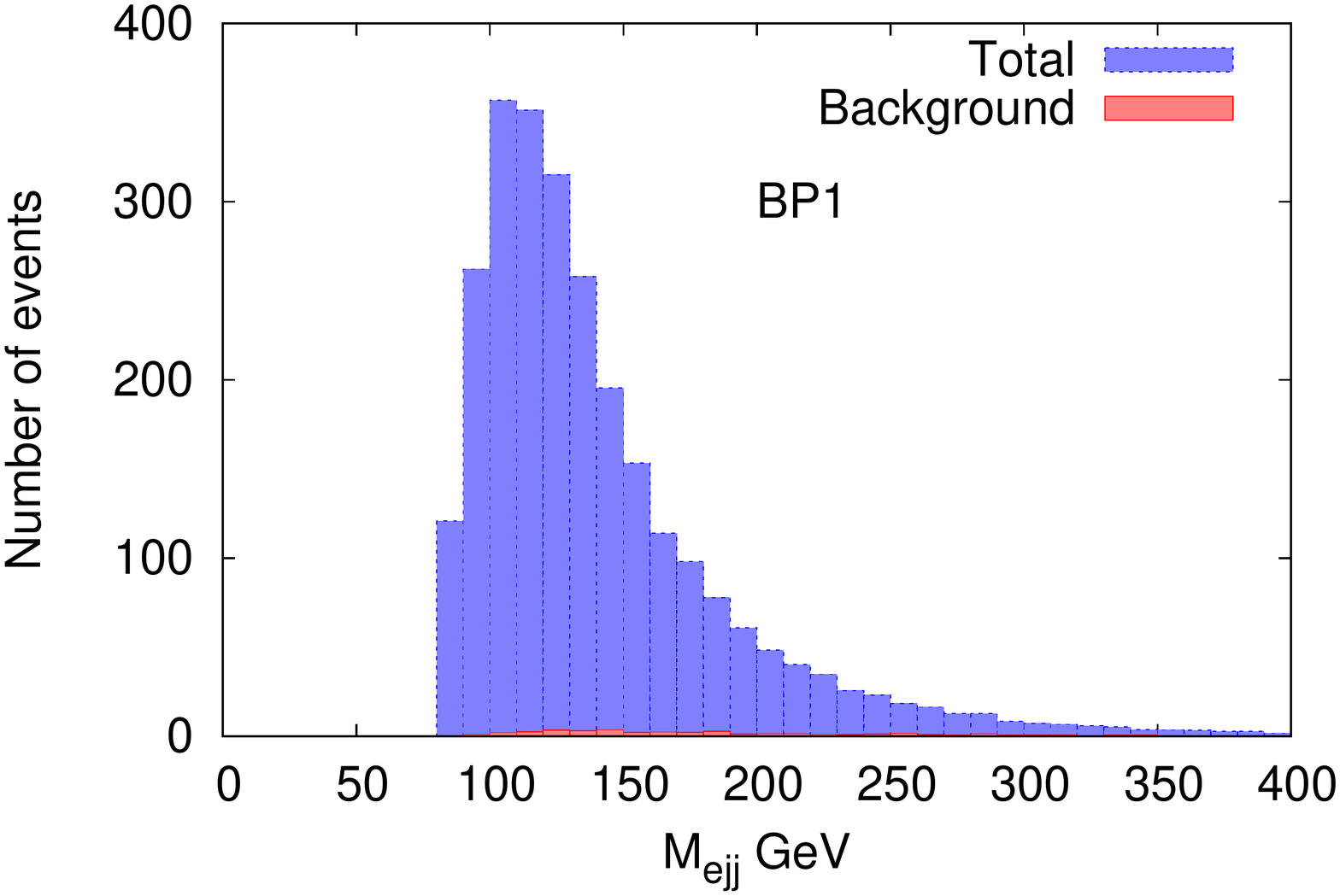}
\caption{ $M_{\ell jj}$ distribution (left) for $n_{\ell}\geq 3 +n_{j}\geq 4(n_b \geq 2)+\ptmiss\geq 100$ GeV and $M_{e jj}$ (right) $n_{\ell}\geq 3 (n_e \geq 2) +n_{j}\geq 4(n_b \geq 2)+\ptmiss\geq 100$ GeV for final state at an integrated luminosity of 50 fb$^{-1}$. The blue graph represents the total number of events and the red corresponds to the SM backgrounds only.}\label{ljj}
\end{center}
\end{figure}

For the $2e$ or $3 \ell$ final states a invariant mass of $\ell jj$ will give the right-handed neutrino mass. In Figure~\ref{ljj}(left) we demonstrate the invariant mass distribution of one electron and two jets coming from $W$. Here we have taken two jets satisfying $|(M_{jj}-M_{W})|\leq 15$  GeV. Thus $M_{ejj}$ reconstruct the decay of the right-handed neutrino, i.e., $N\to e W$. To control the dominant SM background $t\bar{t}$ we have chosen $n_{\ell}\geq 3 +n_{j}\geq 4(n_b \geq 2)+\ptmiss\geq 100$ GeV as final state. The demand of additional jets and $b$-jets reduce the SM backgrounds substantially. Figure~\ref{ljj}(left) shows the total number of events coming from the $\tilde{t}_1$, $\tilde{b}_1$ for BP1 and the SM backgrounds at an integrated luminosity of 50 fb$^{-1}$.  $\tilde{t}_2$, $\tilde{b}_2$ contributions are negligible. We can see that the signal peaked around $\sim 100$ GeV, which is the right-handed neutrino mass, $m_{N}$. Clearly it has more than $60\sigma$ signal significance a 50 fb$^{-1}$ luminosity.

Now if we use the flavour violating decay of the right-handed neutrino and demand that out of the three leptons two of them are electrons, then this suppresses the backgrounds much more than the signal. From Figure~\ref{ljj}(right) we can see
the corresponding invariant mass distribution  of $ejj$
 for the final state of  $n_{\ell}\geq 3 (n_e \geq 2) +n_{j}\geq 4(n_b \geq 2)+\ptmiss\geq 100$ GeV. It is visible that the signal stands out over the backgrounds
much clearly.  Thus the study of third generation squarks decays is very important  which can lead to the information about right-handed neutrino mass  produced in a supersymmetric cascade decay. 
 
\section{Conclusion}

A $U(1)'$ supersymmetric seesaw model with R-parity can be motivated by simultaneous explanation of  
the observed neutrino masses and mixing, the existence of dark matter, and the stabilization of the Higgs boson mass 
assuming TeV-scale SUSY breaking scale. This can induce radiative breaking of the electroweak symmetry as well as the $U(1)'$ 
gauge symmetry.  In this scheme, a right-handed sneutrino $\tilde N_1$ becomes a good thermal dark matter candidate if the extra gaugino $\tilde Z'$ is relatively light. 
The addition of the new decay modes, reduces the experimental lower
bounds of the supersymmetric particles, viz, stops and sbottoms.
Considering stop and sbottom below TeV, we showed that $\tilde{Z}'$ produced from third generation SUSY cascade decays can lead to significant lepton number and flavour violating signatures in final states with multi-lepton accompanied 
by multi-jet (+missing energy) through the decay chain of $\tilde Z' \to N \tilde N_1 \to e^\pm W^\mp \tilde N_1$ if allowed kinematically.  These signatures are going to shed a light not only on the existence of a right-handed neutrino but also a $U(1)'$ model with 
a superpartner of an extra $Z'$ boson. In addition to the conventional same-sign dilepton signal, the flavour differences $2e-2\mu$ and $3e-3\mu$, as well as the $4l$ final state are also promising  channels to look for at the 14 TeV LHC. 

In the case of $\tilde Z'$ being NLSP, early data of LHC14 will be able to probe some optimistic benchmark points. 
If $\tilde{Z}'$ is not the NLSP so that the third generation squark branching fraction is shared with other higgsinos and gauginos, 
much more data are needed to tell us about the model under consideration.

The invariant mass of $\ell jj$ system can successfully reconstruct the
right-handed neutrino mass if it is produced in the supersymmetric
cascade decays. Thus it can shed direct light to the right-handed neutrino spectrum
and it's flavour violating decay directly through supersymmetric
cascade decays.

\section*{Acknowledgement} 

EJC is supported by the NRF grant funded by the Korea government (MSIP) (No.\ 2009-0083526) through 
KNRC at Seoul National University. PB also thanks INFN, Lecce for the support at the finishing part of the project.


\end{document}